\documentclass[journal]{journal}

\usepackage{graphicx}
\usepackage{amsmath}
\usepackage{amssymb}
\usepackage{bm}

\begin{document}

\title{Very-high-precision normalized eigenfunctions for a class of Schr\"odinger type equations}

\author{Amna Noreen$^*$ , K{\aa}re Olaussen$^\dagger$ \\~\IEEEmembership{~Institutt for fysikk, NTNU, Trondheim Norway}\\
\thanks{$*$ Amna.Noreen@ntnu.no} 
\thanks{$\dagger$ Kare.Olaussen@ntnu.no}}

\maketitle
\thispagestyle{empty}

\begin{abstract}
We demonstrate that it is possible to compute wave function
normalization constants for a class of Schr{\"o}dinger type
equations by an algorithm which scales linearly (in the
number of eigenfunction evaluations) with the desired
precision $P$ in decimals.
\end{abstract}

\begin{IEEEkeywords}
Eigenvalue problems;
Bound states;
Trapezoidal rule;
Poisson resummation
\end{IEEEkeywords}

\IEEEpeerreviewmaketitle

\section{Introduction\label{Introduction}}

\fontsize{10}{10pt}\selectfont

\IEEEPARstart{I}{n} a recent paper\cite{AsifEtAl} it was demonstrated that it
is possible to solve some eigenvalue problems of the Schr{\"o}dinger
type to almost arbitrary high precision. The cases presented explicitly
were (i)~the ground state eigenenergy of the anharmonic oscillator,
\begin{equation}
    -\psi'' + x^4\,\psi = \varepsilon\,\psi,
    \label{AnharmonicOscillator}
\end{equation}
which was found to an accuracy of more than $1\,000\,000$
decimal digits, (ii)~the eigenstate number $50\,000$ of
the same equation where the eigenenergy was found to
an accuracy of more than $50\,000$ decimal digits, and
(iii)~the lowest even and odd parity states of the
double-well potential,
\begin{equation}
   -s^2\,\psi_{\pm}'' + (x^2 -1)^2\,\psi_{\pm} = \varepsilon_{\pm}\,\psi_\pm,
\end{equation}
where the two eigenenergies were found to an accuracy of
more than $30\,000$ decimal digits for the case $s=1/50\,000$.

In the latter case the two eigenenergies are degenerate to almost
$29\,000$ decimals, with the difference directly computable by
the WKB method. The $10^{\text{th}}$ order WKB expansion given in
\cite{ZinnJustin} provide an accuracy of about $48$ decimals
for the difference, all of which agrees with the difference
between the two numerical calculated eigenenergies.

Also the eigenenergies of the highly excited states
of equation~(\ref{AnharmonicOscillator}) can
be calculated by the WKB method. The $12^{\text{th}}$ order WKB expansion given
in \cite{Bender_etal} provide an accuracy of about $67$ decimals,
all of which agrees with the numerical result. 

It is certainly difficult to find physical systems where one
need to know eigenenergies to tens of thousands of decimals or more.
However, if one need to compute the wavefunction very accurately
to very high values of $x$ (to f.i.~evaluate matrix elements)
the value depends extremely sensitively on the eigenvalue parameter.
Thus, in our opinion, algorithms for very-high-precision
evaluation of eigenvalues may be useful in combination
with routines for very-high-precision evaluation of
matrix elements and normalization integrals.

In this paper we demonstrate that the latter can be achieved
very simply, with the number of eigenfunction evaluations
growing only linearly with the desired precision $P$ in
decimal digits. Each eigenfunction evaluation will in turn
require a number of high-precision multiplications which
grows linearly with $P$, and each such multiplication
require a CPU time which scales asymptotically between  $P^{1.6}$
and $P\,\log P\,\log\log P$ (depending on which high-precision
multiplication algorithm is used). Thus, the total time to
evaluate one normalization
integral to $P$ decimals precision can be expected to increase
somewhat faster than $P^3$. However, since the eigenfunction
evaluations are independent they can be run in parallel.

The rest of this paper is organized as follows. In section~\ref{Remarks}
we make some general remarks on numerical integration rules.
We base our analysis on the Euler-Maclaurin summation
formula and the Poisson resummation formula. Our conclusion
is that the trapezoidal rule is not only the simplest one
but also the best one. For integrals over finite intervals
some there are endpoint corrections which should be
considered separately; these corrections vanish for our
normalization integrals. In section~\ref{Examples} we consider some
simple example cases similar to our wavefunction
normalization integrals. For an infinite (in principle)
integration range and a fixed number $M$ of integration steps
one must strive for a balance between the error $\varepsilon(h)$
due to using a finite stepsize $h$, and the error $\varepsilon(x_{\text{max}})$
due to covering only a finite integration range. These errors can
be estimated by respectively analysing the Fourier transform $\tilde{f}(p)$ 
of the integrand $f(x)$ as $\vert p \vert \to\infty$ (by the method of
steepest descent), and the asymptotic behaviour of
$f(x)$ as $\vert x \vert\to\infty$. It appears that both analyses
can be extended to wavefunction normalization integrals by use of
the WKB approximation. This extension is done in section~\ref{Wavefunction}.

\section{Remarks on integration formulae\label{Remarks}}

There is no scarcity of numerical integration formulae
in the literature\cite{AbramowitzStegun}. Standard
choices are rules for $I\left\{f\right\} \equiv \int_{a}^{b}\text{d}x\,f(x)$
which reproduces integrals of polynomials below a certain order exactly,
\begin{align}
    I\left\{f\right\}
    &\approx \frac{1}{2}\left(f_{0} + f_{1}\right)h \equiv T(h),\nonumber\\
    I\left\{f\right\}
    &\approx \frac{1}{3}\left( f_{0} + 4 f_1 + f_2\right)h \equiv S(h),\nonumber\\[-1.6ex]
    &\label{IntegrationRules}\\[-1.6ex]
    I\left\{f\right\}
    &\approx \frac{3}{8}\left( f_{0} + 3 f_{1} + 3 f_{2} + f_{3}\right)h \equiv S_{3/8}(h),\nonumber\\
    I\left\{f\right\}
    &\approx \frac{2}{45}\left(7 f_{0} + 32 f_{1} + 12 f_{2} + 32 f_{3} + 7 f_{4}\right)h \equiv B(h),\nonumber
\end{align}
with $h=(b-a)/M$ and $f_m = f(a+m h)$ when there is $M+1$ terms in the integration rule.
These are known respectively as the trapezoidal, Simpson's, Simpson's $\frac{3}{8}$,
and Boole's rule. They are automatically exact for polynomials which are
antisymmetric about the midpoint $\bar{x}=(b-a)/2$. The $M$ independent coefficients
are chosen to give exact results for all symmetric (about $\bar{x}$) polynomials of order below $2M$. However,
as $M$ increases the weight coefficients develop in a suspicious way. To integrate
functions over large intervals to very high precision it seems dubious to extend the
procedure above.

\subsection{Extended Simpson's rule}

An alternative way to handle integration over large intervals is to divide them
into many smaller ones, and apply one of the rules above to each of the
subintervals. We denote this by a bar over the rule, $T(h)\to \bar{T}(h)$ etc.
A rather common choice is to extend Simpson's rule, leading to the formula
\begin{align}
    \int_{a}^{b} \text{d}x\, f(x) \approx \bar{S}(h) \equiv
    \frac{1}{3}&\left( f_{0} + 4 f_{1} + 2 f_{2} + \cdots\right.\nonumber\\[-1.6ex]
    &\label{ExtendedSimpson}\\[-1.6ex]
    &\left.+ 2 f_{M-2} + 4 f_{M-1} + f_{M}\right)h,\nonumber
\end{align}
which looks curious to any member of an equal society. What is wrong with half of
the points? Are they of the wrong gender? Which mysterious force of numerical
error analysis leads to this spontaneous breakdown of translation invariance?
\vspace{3ex}

\hspace{-1.5ex}\includegraphics[width=0.475\textwidth]{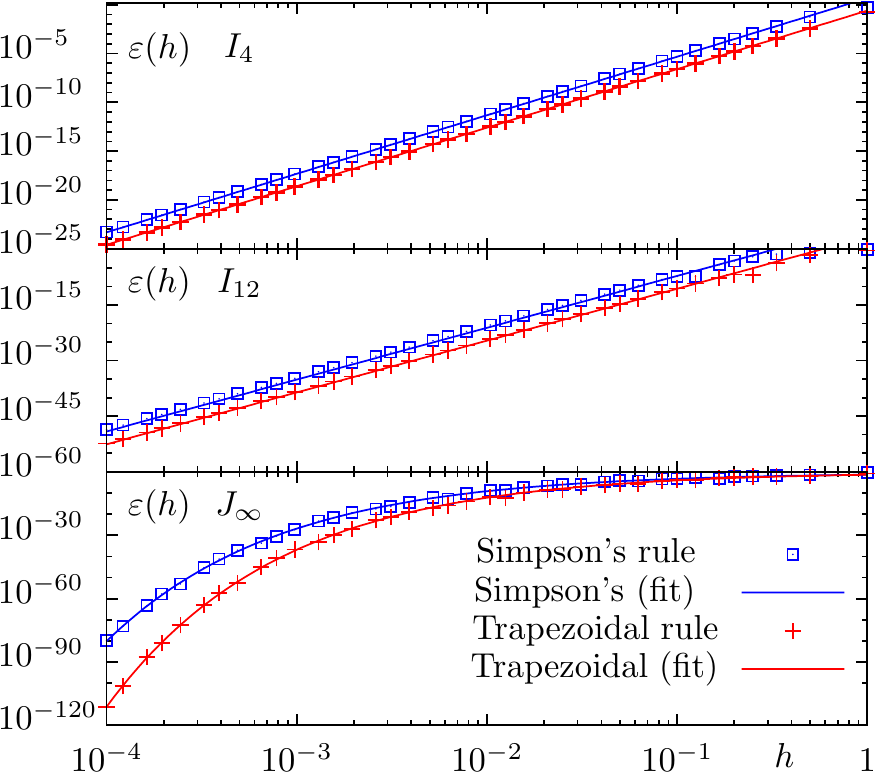}

\noindent
{\fontsize{9}{9pt}\selectfont Fig. 1\ Comparison of accuracy of the extended trapezoidal and Simpson's rules
for some integrands with very smooth boundary behaviour.\/}

\fontsize{10}{10pt}\selectfont
\vspace{3ex}

In our opinion only $\bar{T}(h)$ is a logically sensible procedure.
We claim that this is also the best procedure. To support and exemplify this
claim we have evaluated the integrals below:
\begin{align}
    I_n     &=   \int_0^{1} \text{d}x  \left(1-x^2\right)^n,\\
    J_\infty &=    \int_0^{1} \text{d}x \left[1-\tanh\left(\frac{2x-1}{1-(2x-1)^2}\right) \right],
\end{align}
by the $\bar{T}(h)$ and $\bar{S}(h)=\frac{4}{3}\bar{T}(h)-\frac{1}{3}\bar{T}(2h)$ rules.
The relative errors of the two methods as function
of discretization length $h$ are shown in figure~1. The result is as expected that $\bar{S}$
behave as well as $\bar{T}$ with {\em twice\/} the discretization
length $h$.

\subsection{Euler-Maclaurin summation formula}

The examples in figure 1 are untypical in the sense that the integrands are
very smooth at the endpoints (which is typical for the normalization integrals
we are primarily interested in).
One may use the Euler-Maclaurin summation formula to see which quantity is
actually computed by the various integration rules
{\footnotesize
\begin{align}
    &\left[ \frac{1}{2} f_0 +\sum_{m=1}^{M-1} f_m + \frac{1}{2} f_M\right] h
    = \int_a^b \text{d}x\,f(x)
    + \sum_{k=1}^\infty \frac{B_{2k}}{(2k)!}\Delta^{(2k-1)} + \cdots \nonumber\\ 
    &= \int_a^b \text{d}x\,f(x) + \frac{\Delta^{(1)}}{12}-\frac{\Delta^{(3)}}{720}
    + \frac{\Delta^{(5)}}{30240} -\frac{\Delta^{(7)}}{1209600} + \cdots,\label{EulerMaclaurin}
\end{align}
}
where $\Delta^{(k)} = \left[f^{(k)}(b) - f^{(k)}(a)\right] h^{k+1}$, and the ellipsis in the first
line denote ``non-perturbative'' terms depending on how $f(x)$ varies in the full
integration range.
With $\bar{S}(h)=\frac{4}{3}\bar{T}(h)-\frac{1}{3}\bar{T}(2h)$ one finds the corresponding
quantity for the extended Simpson's rule:
\begin{align}
    &\bar{S}\left\{ f \right\} = \int_{a}^{b} \text{d}x f(x) + 
    \frac{1}{3}\sum^\infty_{k=1}\left(4-2^{2k}\right)\frac{B_{2k}}{(2k)!}\Delta^{(2k-1)}
    + \cdots.
\end{align}
I.e., the $\Delta^{(1)}$-coefficient has been eliminated at the cost of
increasing all the remaining coefficients\footnote{Note that the error terms given for
the ``extended'' rules in reference \cite{AbramowitzStegun} are in disagreement with
our results.}.

A natural way to eliminate
the $\Delta^{(1)}$-coefficient would be to calculate $f'$ explicitly
at the endpoints, or use a numerical approximations like
{\footnotesize
\begin{align*}
    f'(x) &= \phantom{-}\frac{f(x+\delta)  - f(x-\delta)}{2\delta} -\frac{1}{6}f^{(3)}(x)\,\delta^3+\ldots,\\[0.1ex]
    f'(x) &= -\frac{3f(x) - 4 f(x+\delta) + f(x+2\delta)}{2\delta} + \frac{1}{3}f^{(3)}(x)\,\delta^3+\ldots,\\[0.1ex]
    f'(x) &= \phantom{-}\frac{3f(x)-4f(x-\delta)+f(x-2\delta)}{2\delta}+ \frac{1}{3}f^{(3)}(x)\,\delta^3+\ldots,
\end{align*}
}
for sufficiently small $\delta$. One should take $\delta$ somewhat smaller than $h$.
The modi\-fied trapezoidal rule
described in \cite{ModifiedTrapezoidalRule} is generally worse than
the extended Simpson's rule because of an inaccurate approximation of $f'$. However,
there are few reasons to use the same discretization length for computing derivatives
at the endpoints as for computing the bulk contribution to the integral.
To avoid introduction of significant new errors it is sufficient to choose
$\delta < h/2$ in the first formula. By choosing $\delta=h/\sqrt{20}$ in the last two
one also eliminates the $\Delta^{(3)}$-term in equation (\ref{EulerMaclaurin}).

\subsection{Poisson resummation formula}

There exist expressions for the ``non-perturbative'' terms in equation~(\ref{EulerMaclaurin}),
but we find the Poisson resummation formula more clarifying:
Assume $f(x)$ is an entire, integrable function, with Fourier transform
\begin{equation}
   \tilde{f}(p) = \int_{-\infty}^{\infty} \text{d}x\,f(x)\,\text{e}^{\text{i}px}.
\end{equation}
Then one version of the Poisson resummation formula reads
\begin{equation}
    \sum_{m=-\infty}^{\infty} h\,f(mh) = \sum_{k=-\infty}^{\infty} \tilde{f}(\frac{2\pi k}{h}).
\end{equation}
This can be used to estimate the accuracy of the numerical integration formula
\begin{equation}
   \int_{-\infty}^{\infty} \text{d}x\,f(x) = \sum_{m=-\infty}^{\infty} h\,f(m h)
   -\sum_{k\ne 0} \tilde{f}(\frac{2\pi k}{h}).
\end{equation}
When $f(x)$ is an entire function its Fourier transform $\tilde{f}(p)$ will vanish faster that any inverse power
of $p$ as $p\to\infty$. This means that the numerical approximation to the (infinite range) integral of an entire function
will converge {\em very fast\/} towards the exact value as $h \to 0$.

Integrals of $f(x)$ over a finite range $\left[a,b\right]$ can be viewed as integrals of ${g}(x) \equiv \theta(b-x)\,\theta(x-a)\,f(x)$
over an infinite range. But $g(x)$ will usually be discontinuous,
with a fourier transform $\tilde{g}(p)$ which vanishes only algebraically as $p\to\infty$.
The ``perturbative'' endpoint corrections of the Euler-Maclaurin formula can be used
to account for these algebraic terms in a systematic way. But there is no point in making
endpoint corrections beyond the error in  the bulk contribution, the latter being
of magnitude $\tilde{f}(\frac{2\pi}{h}) + \tilde{f}(-\frac{2\pi}{h})$.

\section{Example integrals\label{Examples}}

In this section we will analyze the behaviour of some simple cases which are similar to typical ground state
normalization integrals. 

\subsection{$\bm{\text{e}^{-x^2}}$}

Consider first the ground state of the harmonic oscillator, $f(x) = \text{e}^{-x^2}$, in which case
\begin{equation}
    \tilde{f}(p) = \sqrt{\pi}\,\text{e}^{-p^2/4}.
\end{equation}
Hence we have
\begin{equation}
   \int_{-\infty}^\infty \text{d}x\,\text{e}^{-x^2} = \sum_{m=-\infty}^{\infty} h \text{e}^{-(m h)^2} 
   - \sqrt{4\pi}\,\text{e}^{-\pi^2/h^2} - \cdots.
\end{equation}
In practise we must approximate the infinite sum by a finite one,
\begin{align*}
  \int_{-\infty}^\infty \text{d}x\,\text{e}^{-x^2} =\;
  &h + 2h \sum_{m=1}^{M} \text{e}^{-(mh)^2}\\ 
  +\;&2h \text{e}^{-(M+1)^2 h^2} + \cdots - \sqrt{4\pi}\,\text{e}^{-\pi^2/h^2} - \cdots.
\end{align*}
If we make $h$ too small the first correction term (on the second line above)
becomes too large. If we make $h$ too large the second correction term becomes
too large. I.e., the optimal choice of $h$ occurs approximately when
\(
   \text{e}^{-(M+1)^2 h^2} = \text{e}^{-\pi^2/h^2}
\),
\begin{equation}
    h = \sqrt{\frac{\pi}{M+1}}.
\end{equation}
This leads to an expected error of order
\begin{equation}
    \varepsilon(M) = \text{e}^{-\pi\left(M+1\right)}.
\end{equation}
I.e., to evaluate this integral numerically to $P$ decimals accuracy one must choose
\begin{equation}
   M+1 \ge  \frac{\log 10}{\pi}\,P \approx 0.73\, P.
\end{equation}

\subsection{$\bm{\text{e}^{-x^{2n}}}$}

In more generality consider $f(x) = \text{e}^{-x^{2n}}$ with $n$ a positive integer.
In this case the integral is
\begin{equation}
   K_n \equiv \int_{-\infty}^{\infty} \text{d}x\,\text{e}^{-x^{2n}} 
   = \frac{1}{n} \Gamma\left(\frac{1}{2n}\right).
\end{equation}
We use the saddle point method to estimate its Fourier transform
\begin{equation}
   \tilde{f}(p) = \int_{-\infty}^{\infty} \text{d}x\,\text{e}^{-x^{2n} + \text{i}px} \equiv 
   \int_{-\infty}^{\infty} \text{d}x\,\text{e}^{\phi(x,p)}.
\end{equation}
The saddle point equation becomes
\begin{equation}
     \phi'(x_s,p) = -2n x_s^{2n-1} + \text{i}p =0,
\end{equation}
with $2n-1$ solutions
\begin{equation}
    x_s = \text{e}^{\text{i}\pi(4k+1)/(4n-2)} \left(\frac{p}{2n} \right)^{1/(2n-1)},
    \quad k=0,1,\ldots 2n-2.
\end{equation}
The two most relevant saddle points\footnote{The deformation of the integration path so that
it passes through these saddle points is an interesting exercise\cite{Tafjord}} occur for $k=0$ and $n-1$, at
which the real part of the exponent $\phi(x,p)$ is (when $p=2\pi/h$)
\begin{align}
    \text{Re}\, \phi(x_s,p) &= -(2n-1) \sin\left(\frac{\pi}{4n-2}\right)\,
    \left(\frac{\pi}{n h}\right)^{2n/(2n-1)}
    \nonumber\\
    &\equiv a_n\, h^{-2n/(2n-1)}.
\end{align}
I.e., the leading error due to a finite stepsize $h$ is of
magnitude $\text{e}^{\text{Re}\,\phi(x_s,p)}=\text{e}^{-a_n h^{-2n/(2n-1)}}$ (apart from a prefactor
of less importance). On the other hand, including only the $M$ first terms of the sum
leads to an error of magnitude $\text{e}^{-[(M+1)h]^{2n}}$. For a given $M$ we should
choose $h$ to balance these errors, i.e.
\begin{equation}
    h = b_n\, \left(M+1\right)^{-(1-1/2n)},
    \label{DromedarStepSize}
\end{equation}
with $b_n = \left(\frac{\pi}{n}\right)^{1/2n}\,\left[(2n-1)\sin\left( \frac{\pi}{4n-2}\right) \right]^{(2n-1)/4n^2}$.
This leads to an error of magnitude
\begin{equation}
   \varepsilon(M) = \text{e}^{-c_n(M+1)},
\end{equation}
with $c_n=\frac{\pi}{n}\left[(2n-1)\sin\left(\frac{\pi}{4n-2}\right)\right]^{1-1/2n}$.
I.e., to evaluate the integral numerically to $P$ decimals accuracy one must choose
\begin{equation}
    M+1 \ge \frac{\log 10}{c_n}\, P \approx (0.12+0.467\,n)\,P.
\end{equation}
where the numerical approximation is very good for $n\ge 2$.

\subsection{$\bm{e}^{-(x^2-a^2)^2}$}

Finally consider $f(x)=\text{e}^{-\left(x^2-a^2\right)^2}$.
In this case
\begin{align}
   I(a) &= \int_{-\infty}^\infty \text{d}x\,\text{e}^{-(x^2-a^2)^2}\nonumber\\ &= 
   {a}\,\text{e}^{-\frac{1}{2}a^4}\left[{\textstyle \frac{1}{\sqrt{2}}}\,K_{\frac{1}{4}}({\textstyle \frac{1}{2}}a^4)
     + \pi\, I_{\frac{1}{4}}({\textstyle \frac{1}{2}}a^4)\right]\nonumber\\
   &\mathop{\to}
   \left\{
     \begin{array}{ll}
       2^{-1} \Gamma\left({1}/{4}\right)&\text{as $a\to0$,}\\
       {\sqrt{\pi}}/{a}&\text{as $a\to\infty$.}
     \end{array}
       \right.
\end{align}
The saddle point approximation to the Fourier transform
\begin{equation}
   \tilde{f}(p) = \int_{-\infty}^\infty \text{d}x\,\text{e}^{-(x^2-a^2)^2+\text{i}px}
   \equiv \int_{-\infty}^\infty \text{d}x\,\text{e}^{\phi(x,p)},
\end{equation}
leads to the saddle point equation
\begin{equation}
   \phi'(x_s,p) = -x_s^3 + a^2\,x_s + \text{i} \frac{p}{4} = 0.
\end{equation}
We introduce
\begin{equation}
   x_s = y + \frac{a^2}{3y},
\end{equation}
which leads to a quadratic equation for $y^3$,
\begin{equation}
   y^3 - \text{i}\frac{p}{4} + \frac{a^6}{27 y^3}.
   \label{QuadraticEquation}
\end{equation}
It is convenient to rewrite $p$ so that
\(
    ({p}/{8}) = \left({a}/\!{\sqrt{3}}\right)^3\,\sinh 3\eta
\).
Then the solutions for equation~(\ref{QuadraticEquation}) becomes
\begin{align}
   y^3 &= \text{i}\left( \frac{p}{8} \pm \sqrt{\frac{p^2}{64}+\frac{a^6}{27}} \right)
   = \left(\frac{a}{\sqrt{3}}\text{e}^{\pm (\eta+\text{i}\pi/6)}\right)^3.
\end{align}
I.e.,
\begin{equation*}
   y = \frac{a}{\sqrt{3}}\,\text{e}^{\pm(\eta+\text{i}\pi/6)},\qquad
  \frac{a^2}{3 y} = \frac{a}{\sqrt{3}}\,\text{e}^{\mp(\eta+\text{i}\pi/6)},
\end{equation*}
which in both cases of $\pm$ leads to the solution
\begin{equation}
   x_s = \frac{2 a}{\sqrt{3}}\,\cosh\left(\eta+\text{i}\pi/6\right).
\end{equation}
\vspace{3ex}

\hspace{-2ex}\includegraphics[width = 0.475\textwidth]{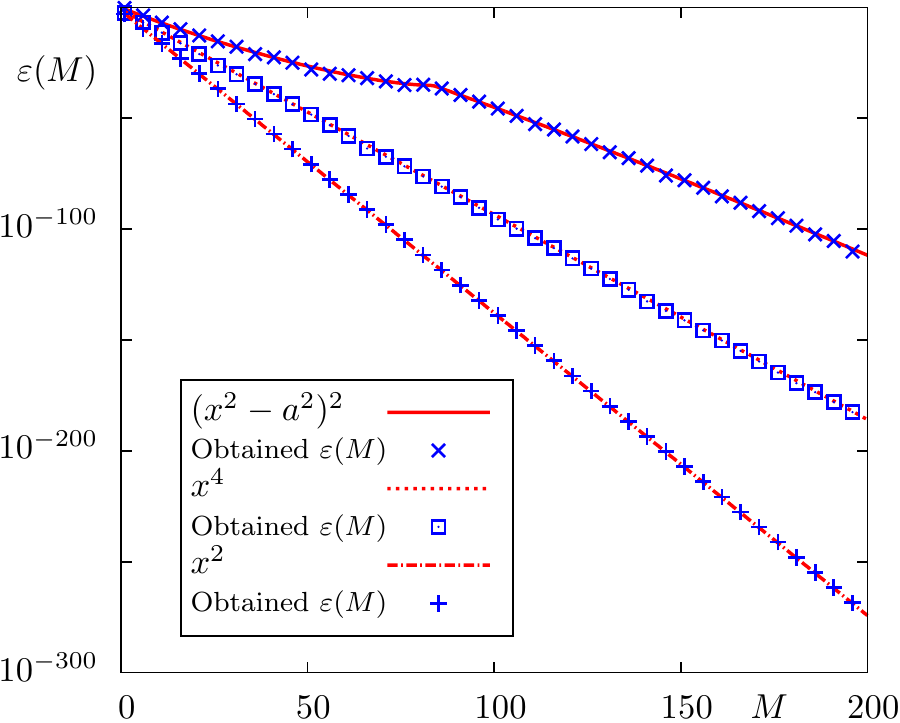}

\noindent
{\fontsize{9}{9pt}\selectfont Fig. 2\ Predicted (lines) and obtained (points) precision as function of $M$, for
the three example integrals discussed in the text. In order from top to bottom:
$\int {\rm d}x\, {\rm e}^{-(x^2-a^2)^2}$, $\int {\rm d}x\,{\rm e}^{-x^4}$,
$\int {\rm d}x\,{\rm e}^{-x^2}$. The
number of function evaluations in each case is $M+1$. As can be seen, the obtained precision
agrees well with the theoretical estimate.
\/}

\fontsize{10}{10pt}\selectfont 
\vspace{3ex}

There are two more ways to take the cube root. One of them leads to an
equally relevant saddle point,
\begin{equation}
   x_s = \frac{2 a}{\sqrt{3}}\,\cosh\left(\eta+\text{i}5\pi/6\right),
\end{equation}
while the last one is irrelevant. The real part of the exponent at
the relevant saddle points is
\begin{equation}
  \text{Re}\, \phi(x_s,\eta)
   = -\frac{4}{3}a^4 \sinh^2\eta\,\cosh 2\eta.
\end{equation}
This provides an error estimate in parametric form: If we choose a finite
stepsize 
\begin{equation}
  h= \frac{\sqrt{27}\,\pi}{4\, a^{3}\sinh 3\eta},
\end{equation}
the corresponding error will be of magnitude
\begin{equation}
  \varepsilon(h) \approx \text{e}^{-\frac{4}{3}a^4 \sinh^2\eta\,\cosh 2\eta}.
\end{equation}
The error caused by summing over only a finite range of $x$-values,
$0\le x_{\text{min}}  \le x \le x_{\text{max}}$,
should be chosen to be of the same magnitude as $\varepsilon(h)$.
I.e., with $s\equiv (4/3)\,a^4 \sinh^2\eta\,\cosh 2\eta$,
\begin{align}
   x_{\text{max}} &= a\,\left(1+\sqrt{s}\right)^{1/2},\\
   x_{\text{min}} &= 
   \left\{
     \begin{array}{ll}
       0&\text{if $s\ge 1$,}\\
       a\,\left(1-\sqrt{s}\right)^{1/2}&\text{otherwise.}
   \end{array}
 \right.
\end{align}
I.e., we must use $M = (x_{\text{max}} - x_{\text{min}})/h$ evaluation steps
in the numerical integration.

\section{Wavefunction normalization integrals\label{Wavefunction}}

Our investigation of the example integrals gives us confidence that
we can obtain a fairly good {\em a priori\/} estimate of the
obtainable precision $\varepsilon(M)$ at a given
number $M$ of discretization steps, at least asymptotically for
large $M$. For this we need to (i) estimate the behaviour of
the Fourier transform, $\widetilde{\psi^2}(p)$, of the integrand at large
$p=2\pi/h$ (to find the obtainable accuracy $\varepsilon(h)$
at a given stepsize $h$), and (ii) estimate how the integrand decays away
from its maxima (to find the required $x$-range of summation for the same
accuracy).

Both quantities can be obtained to reasonable accuracy by use of the
WKB approximation.

\subsection{WKB estimates of wavefunctions for ${x}^{2n}$ potentials}

For large $x$ an estimate of solutions to the eigenvalue problems
\begin{equation}
    -\psi'' +\left(x^{2n} - E \right)\psi = 0,
\end{equation}
can be written in the form\footnote{We ignore the algebraic prefactor $(x^{2n}-E)^{-1/4}$.}
\begin{align*}
    \psi(x) &= \exp{\textstyle \left(-\int_{x_0}^x\,\text{d}t \sqrt{t^{2n}-E}\right)}
    \\
    &\approx C\,\exp{\textstyle \left(-\frac{1}{n+1} x \sqrt{x^{2n}-E}\right)},
\end{align*}
where $x^{2n}_0 = E$, and
\begin{align}
    C &= 
    \exp{\left\{{ \frac{n}{n+1}}\int_{x_0}^{\infty} \text{d}t 
    \frac{E}{\sqrt{t^{2n} - E}}\right\}} \nonumber\\&= 
    \exp\left\{{ \frac{B({\frac{1}{2},\frac{n-1}{2n}})}{2(n+1)}} \,E^{(n+1)/2n}\right\}
    \nonumber\\
    &= \exp\left\{\frac{\pi}{2}\tan\left(\frac{\pi}{2n}\right)\,\left(N+\frac{1}{2}\right)\right\},
    \label{C_expression}
\end{align}
with  $B(x,y) = \Gamma(x)\, \Gamma(y)/\Gamma(x+y)$ the Beta function.
We have evaluated the WKB integral using partial integration,
and let $x\to\infty$ in the remainder. Further, in the last equality
we have assumed $E\equiv E_N$ to be eigenvalue number $N$, and evaluated it
by the WKB approximation.

We use this approximation to estimate the Fourier transform,
\begin{equation}
   \widetilde{\psi^2}(p) \equiv \int \text{d}x\, \psi(x)^2\,\text{e}^{\text{i}px},
\end{equation}
by the saddle point method. The saddle point equation can be written
\begin{equation}
     x_s^{2n} = E - (p/2)^2,
\end{equation}
giving
\begin{align}
    \left|\, \widetilde{\psi^2}(p)\, \right| &\approx C^2\,
    \exp\left({\textstyle \frac{n}{n+1}}\text{Re}\left(\text{i} p x_s\right)\right)
\nonumber\\
    &= C^2\,\exp\left\{-{\textstyle \frac{n}{n+1}}\sin\left({\textstyle \frac{\pi}{2n}}\right)
      \,p\left[ (p/2)^2 -E \right]^{1/2n}\, \right\}.
    \label{FourierEstimate}
\end{align}
Thus, to compute the normalization integral to $P$ decimals precision one must choose $p = 2\pi/h$
so that the right hand side of (\ref{FourierEstimate}) becomes equal to $10^{-P}$ (or smaller).
The value of $h$ is best found numerically. But note that one at least must have
$(p/2)^2 > E$, or
\begin{equation*}
  {h}^{-1} > {\pi}^{-1} E^{1/2}.
\end{equation*}

For large $x$ the normalization integrand behaves like
\begin{align}
     \psi(x)^2 &\approx \exp\left(-2\int_0^x\,\text{d}t \sqrt{t^{2n} - E} \right) \nonumber\\
     &\approx C^2\,\exp\left(-{\textstyle \frac{2}{n+1}} x\sqrt{x^{2n}-E} \right).
     \label{PsiSquareEstimate}
\end{align}
Thus, to compute the normalization integral to $P$ decimals precision one should integrate
to a value $x=x_{\text{max}}$ for which the right hand side of (\ref{PsiSquareEstimate})
becomes less than $10^{-P}$.
The value of $x_{\text{max}}$ is best found numerically. But note that one at least must have
$x_{\text{max}}^{2n} > E$, or
\begin{equation*}
  x_{\text{max}} > E^{1/2n}.
\end{equation*}
This provides a lower limit on the number of integration steps,
\begin{equation}
   M+1 = \frac{x_{\text{max}}}{h} > \frac{1}{\pi}\, E^{(n+1)/2n} \approx  
   \frac{N+\frac{1}{2}}{\int_{-1}^1 \text{d}u \sqrt{1-u^{2n}}}.
   \label{LowerLimit}
\end{equation}
Here we have assumed $E \equiv E_N$ to be eigenvalue number $N$, and used the WKB approximation to
estimate its value. Since $N$ is the number of nodes in the eigenfunction, the
inequality (\ref{LowerLimit}) quite reasonably says that the number of evaluation
points must be larger than the number of oscillations.

\subsection{Ground state wavefunction of ${x}^4$ potential}

Specializing the results of the previous section to the ground state of the
$x^4$-potential, and approximating $E_0$ by zero, one finds from
equation (\ref{FourierEstimate}) that
\begin{equation}
    \left| \widetilde{\psi^2}(p)\right| \approx \text{e}^{-p^{3/2}/3},
\end{equation}
and from equation (\ref{PsiSquareEstimate}) that
\begin{equation}
    \psi(x)^2 \approx \text{e}^{-2 x^{3}/3}.
\end{equation}
From this one deduces that the optimal
choices for $h=2\pi/p$ and $x_{\text{max}}=M h$ are so that
\begin{equation*}
   \text{e}^{-\frac{1}{3}(2\pi/h)^{3/2}} \approx \text{e}^{-\frac{2}{3} (M+1)^3 h^3}.
\end{equation*}
I.e.,
\begin{equation}
     h = 2^{1/9} \pi^{1/3}\,(M+1)^{-2/3} \approx  1.58\,(M+1)^{-2/3}.
\end{equation}
The corresponding estimated error becomes
\begin{equation}
     \varepsilon(M) = \text{e}^{-\frac{1}{3} 2^{4/3}\pi\, (M+1)} \approx \text{e}^{-2.64\,(M+1)}.
     \label{EstimatedError}
\end{equation}

I.e., if one wants to compute the normalization integral to $P$ decimals precision
one has to choose
\begin{equation}
    M+1 = \frac{3 \ln 10}{2^{4/3}\pi}\,P \approx 0.87\, P.
    \label{EstimatedM}
\end{equation}

\hspace{-1.5ex}\includegraphics[width = 0.475\textwidth]{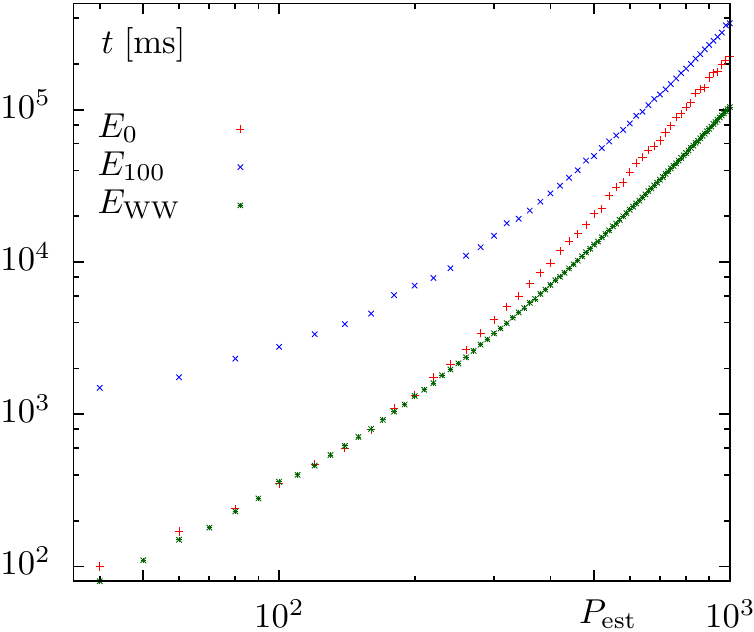}

\noindent
{\fontsize{9}{9pt}\selectfont Fig. 3\ This figure displays the computer time (measured in milliseconds)
required to compute some normalization integrals to $P$ decimals estimated precision.
The cases considered are (i) the ground state wavefunction of the $x^4$-potential
($E_0$), (ii) the $100^{\text{th}}$ excited state wavefunction of the $x^4$-potential
($E_{100}$), and (iii) the ground state wavefunction of the $(x^2 -1)^2$-potential
($E_{\text{WW}}$) with $s={1}/{100}$.
\/}

\fontsize{10}{10pt}\selectfont 

\subsection{Excited state wavefunctions of $x^4$ potential}

The main effect of having highly excited states $E_N$ lies in the extra factors
$C^2$ in the error estimates, with $C$ found in equation (\ref{C_expression}).
Clearly $C$ becomes large for large $N$. To account for this factor we must
replace $10^{-P}$ by $C^{-2}\,10^{-P}$ in the error analysis. I.e., make the replacement
\begin{equation}
   P\,\log 10 \rightarrow P\,\log 10 + 
   \pi \tan \left(\frac{\pi}{2n}\right) \left( N+ \frac{1}{2}\right)
\end{equation}
in the expressions like (\ref{EstimatedM}). There are also $E$-dependent
corrections to the remainding factors, but they are best treated
numerically.

\subsection{Ground state wavefunction of ${(x^2-1)^2}$ potential}

We next consider the lowest even parity eigenstate of
\begin{equation}
   -s^2\, \psi'' + (x^2-1)^2\,\psi = \varepsilon\,\psi,
\end{equation}
with $s$ is small and positive. By WKB analysis one finds that
the solution behaves like
\begin{equation}
  \psi(x) \sim \text{e}^{-(x-1)^2(x+2)/3s},
  \label{DoubleWellGroundState}
\end{equation}
(up to an algebraic prefactor) in the region of interest
($\text{Re} x > 0$, $\vert (x-1)^2(x+2)/3s \vert \gg 1$).
We again estimate Fourier transform of $\psi(x)^2$ by the saddle point
method, assuming the relevant saddle point to be in a region where
the approximation (\ref{DoubleWellGroundState}) is valid. The saddle point
equation,
\begin{equation}
   \phi'(x,p) = -\frac{\text{d}}{\text{d}x} \left[\frac{2}{3s} (x-1)^2(x+2) -\text{i}px\right] = 0,
\end{equation}
has a solution consistent with this assumption,
\begin{equation}
    x_s = (1+\text{i}\frac{ps}{2})^{1/2}.
\end{equation}
The corresponding exponent is
\begin{equation}
   \phi(x_s,p) =\frac{4}{3s}\left[(1+\text{i}ps/2)^{3/2} - 1 \right] 
   \approx -\frac{1}{3s}(1-\text{i}) (ps)^{3/2},
   \label{WWexponent}
\end{equation}
where the last equality is valid for $ps \gg 1$.
To evaluate the integral to $P$ decimals precision one must
choose $p=2\pi/h$ so that $\phi(x_s,p) = -P\,\log 10$ (or smaller).
For very large $P$ one may use the last approximation in
equation~(\ref{WWexponent}) to find
\begin{equation}
   h \approx \frac{2\pi}{\left(3\log 10\right)^{2/3}}\,s^{1/3}\,P^{-2/3} \approx 1.73 \,s^{1/3}\,P^{-2/3},
\end{equation}
otherwise it is simplest to determine $h$ numerically. Further, to evaluate
the integral to $P$ decimals precision one must add contributions
from the (positive) $x$-range where
\begin{equation}
   \psi(x)^2 \approx \text{e}^{-2(x-1)^2(x+2)/3s} > 10^{-P}.
\end{equation}
I.e., we must take $x_{\text{max}}$ to be the largest solution to
the equation $(x-1)^2(x+2)= (3/2) \log 10\, sP$. For very large $P$
the solution becomes
\begin{equation}
    x_{\text{max}} \approx \left({\textstyle \frac{3}{2}}\log 10 \right)^{1/3}\, s^{1/3}\, P^{1/3}
    \approx 1.51\, s^{1/3}\, P^{1/3}.
\end{equation}
Hence the number of required function evaluations again grow asymptotically for large $P$ like
\begin{equation}
  M+1 = \frac{x_{\text{max}}}{h} \approx  0.87\,P,
\end{equation}
cf. equation (\ref{EstimatedM}).

\vspace{3ex}

\hspace{-1.5ex}\includegraphics[width = 0.475\textwidth]{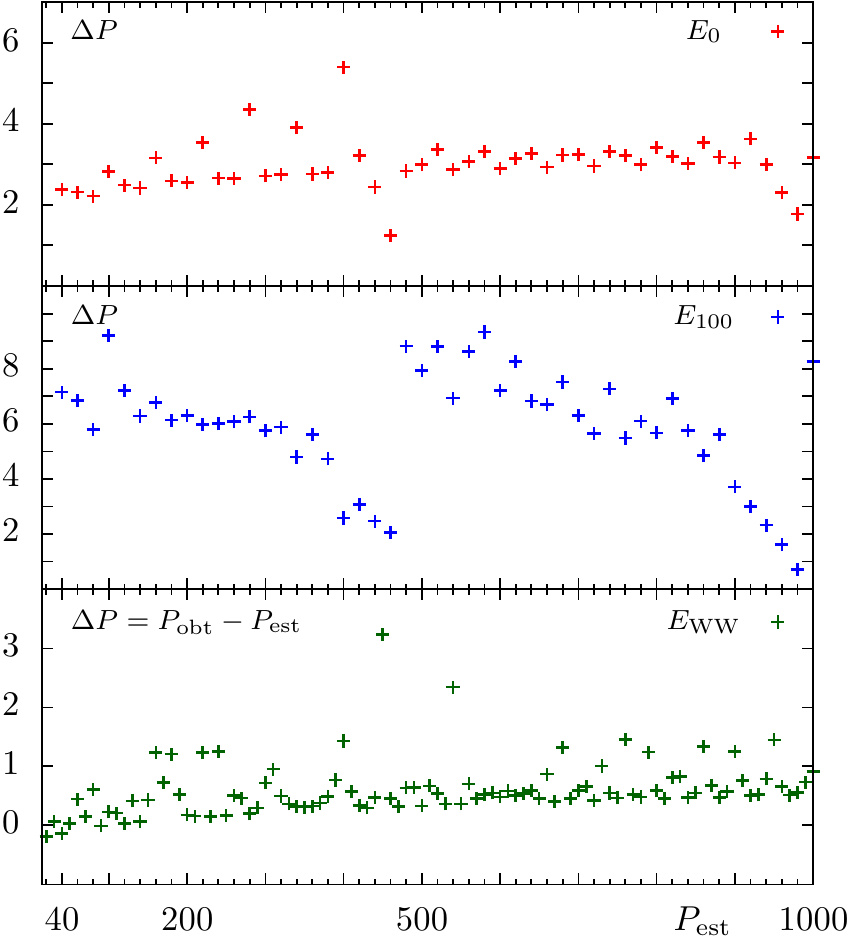}
{\fontsize{9}{9pt}\selectfont Fig. 4\ These figures display the difference between the actually
ob\-tained precision $P_{\text{obt}}$  and the estimated one $P_{\text{est}}$,
for some nor\-mal\-i\-zation integrals, with precisions measured in decimal digits.
The cases considered are (i) the ground state wavefunction of the $x^4$-potential
($E_0$), (ii) the $100^{\text{th}}$ excited state wavefunction of the $x^4$-potential
($E_{100}$), and (iii) the ground state wavefunction of the $(x^2 -1)^2$-potential
($E_{\text{WW}}$) with $s={1}/{100}$.
\/}

\fontsize{10}{10pt}\selectfont 

\newpage
\section*{Acknowledgement}
\label{acknowledgement}
This work was supported in part by the Higher Education Commision of Pakistan (HEC)
and the ''SM{\AA}FORSK'' program of the Research Council of Norway. The calculations
were performed using the CLN\cite{CLN} high-precision numerical library. 
\ifCLASSOPTIONcaptionsoff
  \newpage
\fi

\end{document}